\documentclass{aa}

\usepackage{graphicx}
\usepackage{subcaption}
\usepackage{color}

\usepackage{txfonts}
\usepackage{multirow}
\usepackage[colorlinks=true,linkcolor=blue,filecolor=blue,citecolor=blue,urlcolor=blue]{hyperref}

\begin{document}

   \title{Phase-resolved X-ray spectroscopy of the Vela pulsar with \textit{NICER} observations}

   \author{E. Zubieta
          \inst{1,2}\fnmsep\thanks{E-mail: ezubieta@iar.unlp.edu.ar}
          \and
          F. Garc\'ia\inst{1,2}
          \and
          S. del Palacio\inst{1,3}
          \and
          J. A. Combi\inst{1,2}
          \and
          F. A. Fogantini\inst{4,5}
          }

   \institute{Instituto Argentino de Radioastronom\'ia (CCT La Plata, CONICET; CICPBA; UNLP),
              C.C.5, (1894) Villa Elisa, Buenos Aires, Argentina
         \and
             Facultad de Ciencias Astron\'omicas y Geof\'{\i}sicas, Universidad Nacional de La Plata, Paseo del Bosque, B1900FWA La Plata, Argentina
         \and
            Department of Space, Earth and Environment, Chalmers University of Technology, SE-412 96 Gothenburg, Sweden
        \and
            The Catholic University of America, 620 Michigan Ave., N.E. Washington, DC 20064, USA 
        \and
            NASA Goddard Space Flight Center, Greenbelt, MD 20771, USA
}
   \date{Received ...; accepted ...}

  \abstract
   {PSR J0835$-$4510 (a.k.a. the Vela pulsar) is the best studied pulsar in the Southern hemisphere given its high radio brightness and large glitch activity. However, its X-ray brightness is comparatively lower, making the study of its X-ray profile difficult.}
   {We aim to investigate the X-ray emission of the Vela pulsar coming from the surface of the neutron star to infer the properties of its hotspots and assess whether they share a common spectral shape. We also aim to prove whether the 2021 glitch affected the hotspots. % by comparing the pre-glitch and post-glitch pulse profiles.
   }
   {We used ephemeris obtained from high-cadence radio observations with the radio telescopes of the Argentine Institute of Radio Astronomy to combine all the \textit{NICER} observations of the Vela pulsar after its 2019 glitch and before its 2024 glitch. We obtained a folded pulse profile for three different energy ranges and subsequently studied their phase-resolved energy spectra. We additionally analyzed a \textit{Chandra} observation to quantify the pulsar's contribution to the total soft X-ray emission detected by \textit{NICER}, which cannot spatially resolve the pulsar from the surrounding pulsar wind nebula (PWN).} 
   {The pulsar contribution to the total observed emission represents less than 40–45\% of the flux below 0.9 keV, dropping to 3–4\% in the 0.9--2.0~keV range and further down to $\sim$2\% above 2 keV, where the PWN fully dominates. We find that the X-ray folded profile of the Vela pulsar was not measurably perturbed by the 2021 glitch. We studied separately seven regions of the pulse profile. We found that all of them are well modeled by one or two blackbodies. The temperatures of these hotspots are between $0.79(7) \times 10^{6}\,\mathrm{K} \leq T \leq 4.8^{+0.8}_{-0.6} \times 10^{6}\,\mathrm{K}$, while their sizes vary between $0.048^{+0.016}_{-0.009}\,\mathrm{km} \leq R \leq 3.9^{+0.9}_{-0.6}\,\mathrm{km}$, and their luminosities between $5\times10^{30}~\mathrm{erg}~\mathrm{s^{-1}} \lesssim L < 1\times10^{32}~\mathrm{erg}~\mathrm{s^{-1}}$.}
   {}


   \maketitle
%
%-------------------------------------------------------------------

\section{Introduction}

The Vela pulsar has particular properties that make it the most studied pulsar from the southern hemisphere. It is relatively young ($\sim 11~\mathrm{kyr}$), nearby ($\sim290~\mathrm{pc}$) \citep{2003ApJ...596.1137D} and very bright in the radio band, with a flux of $\mathrm{S_{1400}}=1050~\mathrm{mJy}$ \citep{2005AJ....129.1993M}. It has a rotation period of $P\sim89~\mathrm{ms}$ and a period derivative of $\dot P=1.24 \times10^{-13}~\mathrm{s}~\mathrm{s^{-1}}$. It was the first pulsar known to have glitches \citep{1969Natur.222..228R}, and is the only one that presents glitches quasi-periodically \citep{2017MNRAS.469.4183A, 2022MNRAS.510.4049B,2020MNRAS.496.2506G, 2024PASA...41..102G, 2026PASA...43...25G}, with a strong glitch every $\sim 3~\mathrm{yr}$. 

In addition, the pulsar can be detected in different wavelengths \citep{2015MNRAS.449.3827K}. Pulsed emission is strongly detected in the radio band \citep[e.g.]{2018Natur.556..219P, 2025A&A...698A..72Z}. It is the brightest persistent source in the GeV band, which makes it one of the most studied pulsars in the $\gamma$-ray band \citep{1975ApJ...200L..79T, 2009ApJ...696.1084A, 2022MNRAS.511..425G}. Moreover, its pulsed emission is also detected in the optical \citep{2019MNRAS.482..175S}, ultra-violet \citep{2005ApJ...627..383R}, near-infrared \citep{2003A&A...406..645S} and X-ray bands \citep{1993Natur.361..136O}. The pulsar is surrounded by its wind nebula (Vela-X PWN), which is also strongly detected from radio to $\gamma$ rays \citep{2003MNRAS.343..116D, 2010ApJ...713..146A,2018A&A...617A..78T,2025ApJ...988..200L}.

In the soft X-ray band, the first detailed study of the Vela system was carried out with the IPC and HRI instruments aboard the \textit{Einstein} satellite. However, they did not detect confidently pulsed emission from the object associated with the Vela pulsar \citep{1985ApJ...299..828H}. Later, \cite{1993Natur.361..136O} detected unambiguously pulsed emission from the pulsar, with observations from the Roentgen Satellite (\textit{ROSAT}) in the 0.1--2.4 keV band. This emission was confirmed by \cite{2000ApJ...536..948S}, revealing a profile with one broad and two narrow peaks. Next, the high resolution of the \textit{Chandra} and \textit{XMM–Newton} observatories made it possible to separate the pulsar from the PWN and study its spectrum. The soft part of the X-ray spectrum is usually described as a power law plus a component of thermal emission or a magnetized atmosphere \citep{2001ApJ...552L.129P}.

Glitches have been associated with changes in the pulse profile in some pulsars \citep{2018MNRAS.478L..24K, 2022ApJ...931..103L, 2024MNRAS.533.4274L}. In the Vela pulsar, \cite{Palfreyman+2016, 2018Natur.556..219P} reported changes in the single-pulse radio emission during a glitch, suggesting that the magnetospheric configuration can be transiently affected. Moreover, persistent changes in the radio pulse profile following glitches have been detected in young pulsars \citep{2018MNRAS.478L..24K, 2022ApJ...931..103L, 2025A&A...694A.124Z}.

The X-ray pulse profile of the Vela pulsar is a unique window to investigate the physics of emission mechanisms and magnetospheric dynamics given that it presents both thermal and non-thermal emission. The analysis of the pulse profile shape can probe the emission regions and magnetic field structure. Therefore, by combining pulse shape study and spectroscopy, it is possible to disentangle the thermal and non-thermal components for each emitting region. In this context, it is also of interest to investigate whether a glitch event can produce detectable changes in the X-ray pulse profile, probing potential reorganizations of the surface emission regions.

In order to obtain a high signal-to-noise X-ray profile with \textit{Neutron star Interior Composition Explorer} (\textit{NICER}) observations, it is essential to combine different observations with accurate ephemeris of the pulsar. The Vela pulsar is monitored with the antennas of the Argentine Institute of Radioastronomy (IAR) with up to a daily cadence \citep{2020A&A...633A..84G, 2025A&A...698A..72Z}, which allows us to obtain high-precision ephemeris of the pulsar. In this work, we used the radio observations to combine 321 observations of \textit{NICER} and studied the phase-resolved spectrum for each detected emission region. With the aim of better characterizing the emission of the folded pulse profile obtained with \textit{NICER} observations, we also use a \textit{Chandra} observation to characterize the non-pulsed emission of the Vela pulsar.

The paper is organized as follows: in Sect.~\ref{sec: obs} we describe the process for obtaining and analyzing the observations from IAR, \textit{NICER} and \textit{Chandra}. In Sect.~\ref{sec: results} we present the analysis of the \textit{Chandra} data, the ephemeris obtained with IAR observations, and the pulse profile and phase-resolved spectra obtained with \textit{NICER} observations. In Sect.~\ref{sec: discussion} we discuss the implications of our results.

%===============================================================================================

\section{Observations and data reduction}\label{sec: obs}

\subsection{Radio observations}

The IAR, located near the city of La Plata in Argentina, hosts two 30-m single-dish radio telescopes positioned 120~m apart. The two antennas, "Carlos M. Varsavsky" and "Esteban Bajaja", are also referred to as A1 and A2, respectively. Both can follow sources in the southern hemisphere for up to $3.6~\mathrm{h}$ per day. Observations with A1 have
one circular polarization, and a bandwidth of 112~MHz at a central frequency of 1400 MHz, while observations with A2 have
a bandwidth of 56 MHz at a central frequency of 1428 MHz in two circular polarizations \citep{2020A&A...633A..84G}.

Since 2019, the Pulsar Monitoring in Argentina (PuMA\footnote{https://puma.iar.unlp.edu.ar/}) group uses the antennas to observe with high cadence a set of bright pulsars of the southern hemisphere. Particularly, the Vela pulsar is one of the main targets of the observations of the PuMA group. The group reported Vela's 2021 \citep{2021ATel14806....1S} and 2024 \citep{2024ATel16608....1Z} glitches, which were thoroughly analyzed in \cite{2023MNRAS.521.4504Z,2024A&A...689A.191Z, 2025A&A...698A..72Z}.

To construct a single phase-connected radio ephemeris spanning the entire baseline of interest, we used radio timing observations from IAR between February 8, 2019 (MJD 58522, seven days after the 2019 glitch) and April 28, 2024 (MJD 60428, our last observation before the 2024 glitch). This window is bounded by the two glitches neighboring the July 2021 event (MJD 59417) and coincides with a dedicated NICER monitoring campaign of the Vela pulsar. We did not extend the radio baseline earlier than the 2019 glitch, since we lack radio data before it and this window already covers most of the NICER monitoring campaign. We stopped at the 2024 glitch because subsequent NICER coverage is negligible.

During this period, at the IAR we extensively monitored the Vela pulsar. We collected 466 observations with A1 and 737 with A2. We used these observations to obtain an accurate timing solution during the selected data span, which is necessary to merge the X-ray observations. We processed the radio observations with PRESTO \citep{2003ApJ...589..911R,2011ascl.soft07017R}. We removed radio-frequency interferences (RFIs) and folded the data using the tasks \textsc{rfifind} and \textsc{prepfold}, respectively. Finally, we obtained the times of arrival (TOAs) of the pulses by cross-correlating each folded profile with a template with the \textsc{pat} task in PSRCHIVE \citep{2004PASA...21..302H}. We obtained the template applying \textsc{psrsmooth} (also from PSRCHIVE) to a high signal-to-noise ratio (S/N) profile.

We used the high-cadence radio observations from the IAR to obtain accurate ephemeris of the Vela pulsar. We fitted the TOAs to a Taylor series with TEMPO2 \citep{2006MNRAS.369..655H} to characterize the rotation of the pulsar:

\begin{equation}\label{eq:timing-model}
    \phi(t)=\phi+\nu(t-t_0)+\frac{1}{2}\dot{\nu}(t-t_0)^2+\frac{1}{6}\Ddot{\nu}(t-t_0)^3,
\end{equation}

where $\nu$, $\dot\nu,$ and $\ddot\nu$ are the rotation frequency of the pulsar, and its first and second derivatives, respectively, with $t_0$ their reference epoch. The series can be expanded to higher orders if necessary.

In case of a glitch, the glitch-induced changes in the phase  can be included in the timing model as \citep{2013MNRAS.429..688Y}:

\begin{multline}\label{eq:glitch-model}
    \phi_\mathrm{g}(t) = \Delta \phi + \Delta \nu_\mathrm{p} (t-t_\mathrm{g}) + \frac{1}{2} \Delta \dot{\nu}_\mathrm{p} (t-t_\mathrm{g})^2 + \\ 
    \frac{1}{6} \Delta \Ddot{\nu}(t-t_\mathrm{g})^3 + \left[1-\exp{\left(-\frac{t-t_\mathrm{g}}{\tau_\mathrm{d}}\right)} \right]\Delta \nu_\mathrm{d} \, \tau_\mathrm{d}.
\end{multline}
\noindent Here, $\Delta \phi$ is used to counteract uncertainty in the epoch of the glitch $t_\mathrm{g}$. Next, $\Delta \nu_\mathrm{p}$, $\Delta \dot{\nu}_\mathrm{p}$, and $\Delta \Ddot{\nu}$ are the permanent shifts in $\nu$, $\dot\nu,$ and $\Ddot{\nu}$, respectively, with respect to the preglitch solution. Finally, $\Delta \nu_\mathrm{d}$ is the temporary increase in frequency that will recover exponentially after $\tau_\mathrm{d}$ days. 

\subsection{Chandra observations}
 
The field of the Vela pulsar was observed with the \textit{Chandra} ACIS instrument in June 2022 for 61 ks (ObsID = 24711), which is the closest observation in time to the 2021 glitch. We calibrated the \textit{Chandra} observation using the CIAO software package (version 4.17) and CALDB (version 4.12.2) by running the \texttt{chandra\_repro} task. 

Fig.~\ref{fig: chandra_image} shows a three-color composite image of the \textit{Chandra} observation in three energy bands: soft (0.5--1.5 keV) in red, medium (1.5--2.5 keV) in green, and hard (2.5--8.0 keV) in blue. 

We extracted spectra from the \textit{Chandra} observation for multiple regions indicated in Fig.~\ref{fig: chandra_image} using the CIAO task \texttt{specextract}. The extraction regions consist of a 2.5'' circle centered on the Vela pulsar, and a large co-aligned circle of 1' radius encompassing the PWN surrounding the Vela Pulsar, as well as two circular regions of 2.5'' radius in the innermost parts of the PWN, close to the Vela pulsar. The spectra were binned to 16 counts per bin, and the \texttt{correctpsf} option was enabled for the smallest region centered on the Vela PSR, as most of its emission originates from a point-like source.

\subsection{NICER observations}

\textit{NICER} was launched on June 3, 2017, and is attached to the International Space Station. Its large effective area in soft X-rays makes it an optimal tool for studying pulsar emission. It consists of 56 co-aligned detectors that operate in the range of 0.2--12 keV with a time resolution of $0.1~\mathrm{\mu s}$.  For our analysis, we only consider events above 0.25 keV \citep{2024ApJ...971L..20C}.

We selected all the observations after the 2019 Vela glitch and before the 2024 glitch, which represents $63\%$ of the total NICER dataset of the Vela pulsar. It coincides with high-cadence radio monitoring at the IAR, and it encompasses the 2021 glitch, allowing a direct comparison of the pulse profile before and after the event. We used 321 observations between February 2019 and April 2024 (MJD 58518--60426) that, in total, correspond to a total exposure time of $530.8~\mathrm{ks}$. The duration of individual observations ranges between $12~\mathrm{s}$ and $10453~\mathrm{s}$. We analyzed the data with the \textit{NICER} software version \textsc{2024-08-18\_V013} included in the HEAsoft distribution \textsc{12Aug2024\_V6.34}. 
For each observation, we started the analysis with the \textsc{nicerl2} calibration task. Next, we used the \textsc{photonphase} task from the PINT\footnote{https://github.com/nanograv/PINT} software to apply barycentric corrections and fold the observations with the ephemeris obtained from the radio observations. We merged the observations with the \textsc{niextract-events} task of the \textit{NICER} software. We found the optimal energy range with the \textsc{nioptcuts.py} task of \textit{NICERsoft}\footnote{\url{https://github.com/paulray/NICERsoft}}, that performs an H-test to look for the best energy range in the whole NICER energy range.

We also studied the phase-resolved spectra of the merged observation. We used the task \textsc{nicerl3} without background to obtain the spectrum of each region. We used as background a region that we defined as the off-pulse emission. We carried this out by filtering in phase using a GTI file created with the \textsc{maketime} task. In this way, we obtained the spectrum corresponding to the phases of interest of the integrated profile. We studied the spectrum in the range 0.25--12~keV.

\section{Results}\label{sec: results}

\subsection{Analysis of the \textit{Chandra} data}\label{sec: chandra}

\begin{figure}%[h!]
    \includegraphics[width=\linewidth]{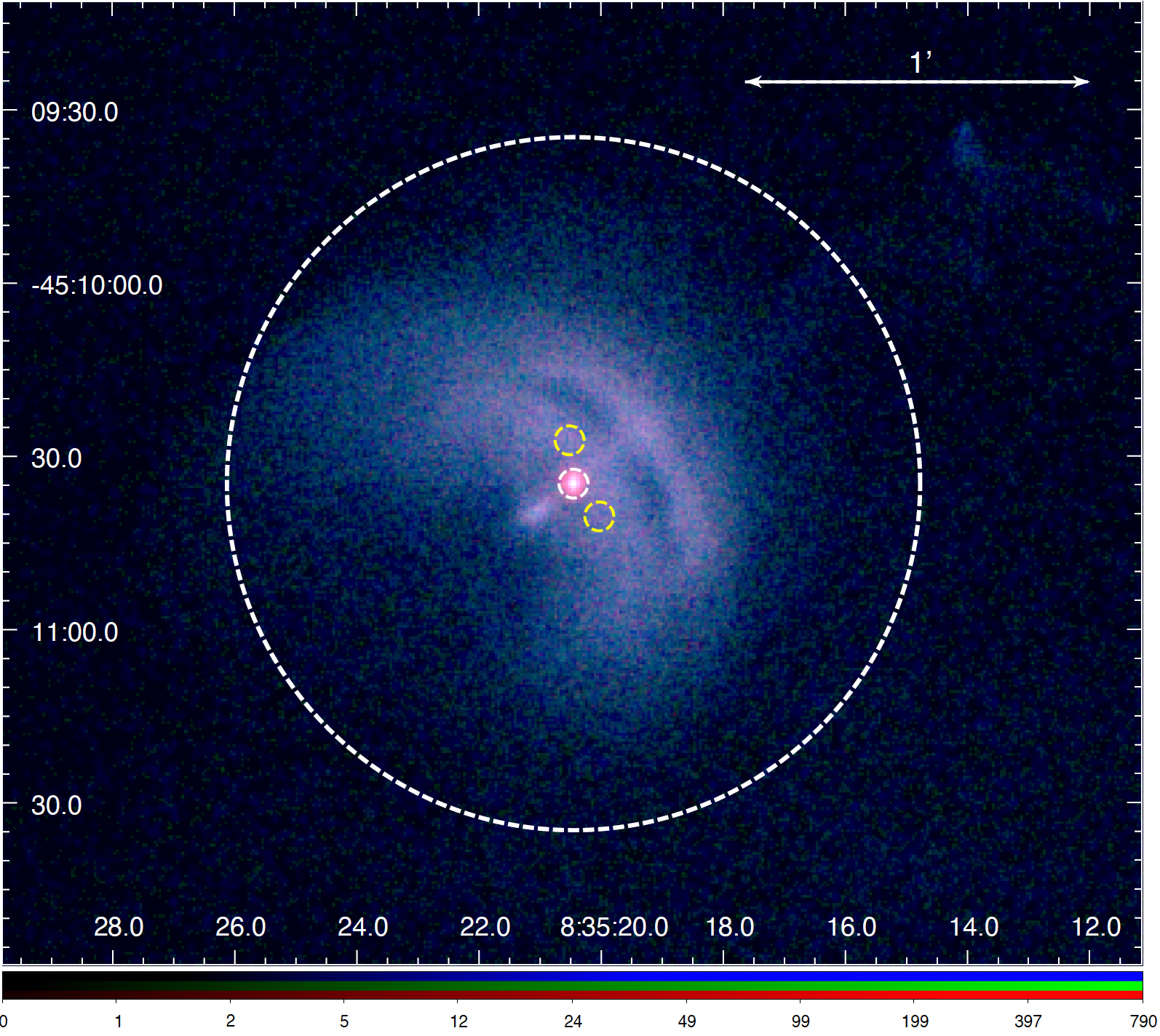}
    \caption{\textit{Chandra} image of the Vela pulsar and its pulsar wind nebula, corresponding to the ObsID 24711 (June 2022). Red color represents the soft (0.5--1.5 keV), green color represents the medium (1.5--2.5 keV), and blue color represents the hard (2.5--8.0 keV) energy ranges. The small yellow regions encircling the Vela PSR and the surrounding PWN have radii of 2.5'', while the large circle encompassing the full emission has a radius of 1', similar to the \textit{NICER} FoV.}
    \label{fig: chandra_image}
\end{figure}

In Fig.~\ref{fig: chandra_image} we present the \textit{Chandra} X-ray image of the Vela pulsar and its pulsar wind nebula, revealing a highly structured morphology in the immediate vicinity of the pulsar. The emission is dominated by a bright inner nebula, from which a pair of collimated, axial jets emerge. Oriented perpendicularly to these jets, a series of diffuse arcs, interpreted as the termination shock of the relativistic pulsar wind, delineate a toroidal geometry. This jet–torus framework is centrally embedded within a larger, asymmetric X-ray synchrotron nebula. 

\begin{figure}%[h!]
    \includegraphics[width=\linewidth, trim={0.7cm 1.6cm 2.9cm 2.7cm}, clip]{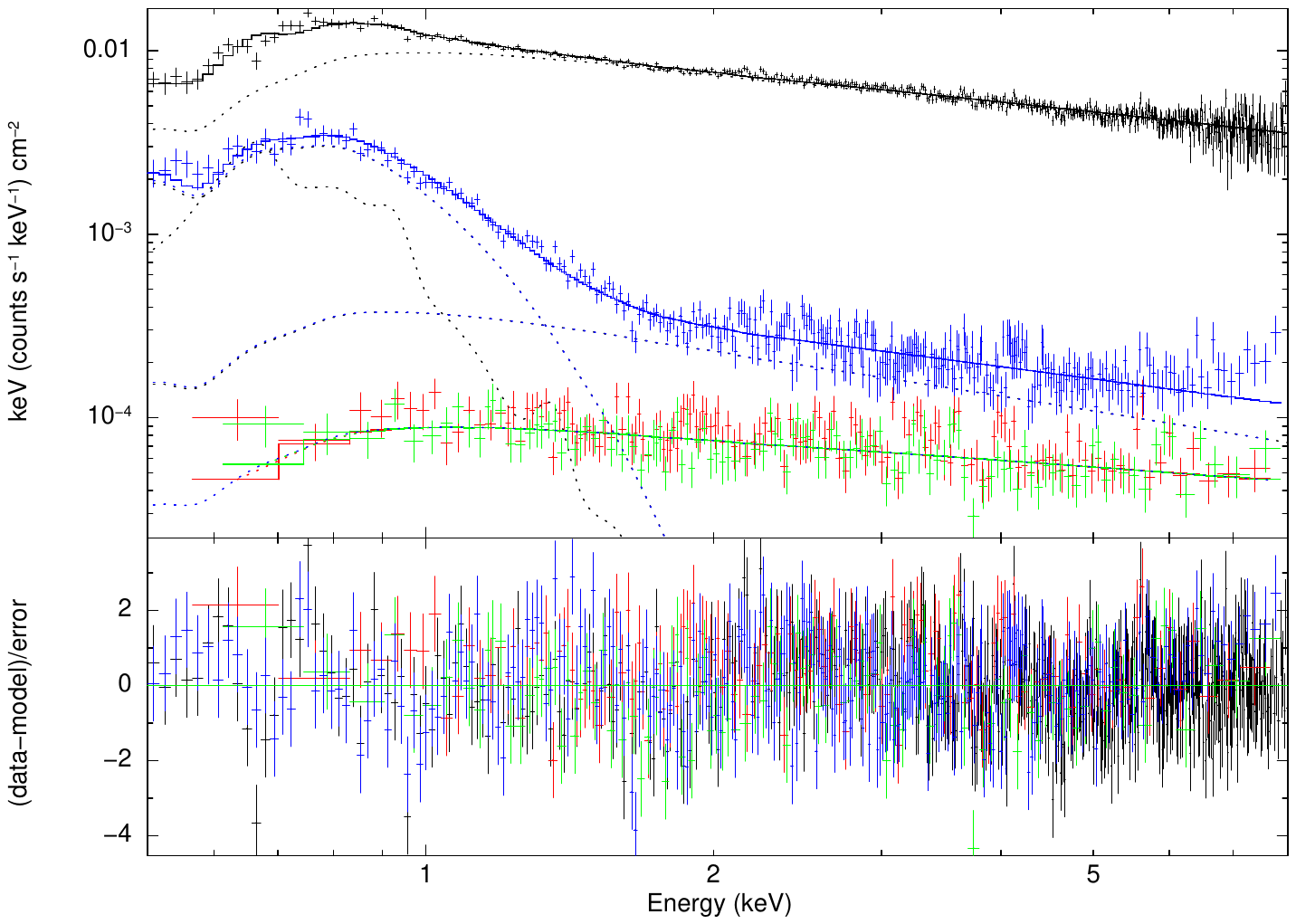}
    \caption{\textit{Chandra} spectra of the Vela pulsar and its pulsar wind nebula, corresponding to ObsID 24711 (June 2022). Red and green spectra correspond to the PWN emission from the $2.5\arcsec$ circles close to the central PSR marked in Fig.~\ref{fig: chandra_image}. The blue data correspond to the spectrum extracted from a circular region of $2.5\arcsec$ centered at the Vela PSR. Black data correspond to the spectrum obtained from the large circle encompassing the full $1\arcmin$ emission, mimicking the \textit{NICER} FoV. The bottom panel shows the residuals of the best-fitting models (see text for details). }
    \label{fig: chandra_spectrum}
\end{figure}

In such Fig.~\ref{fig: chandra_image}, we also show the selected X-ray spectral extraction regions indicated with yellow and white dashed circles. The \textit{Chandra} spectra are used to estimate the relative contribution of the pulsar spectrum to the full spectrum of the Vela pulsar plus the Vela PWN that is observed by the \textit{NICER} instrument, given its comparatively poor angular resolution (of the order of 5'). For this purpose, we first selected a circle of 2.5'' radius centered at the Vela pulsar, to obtain the spectrum of the Vela pulsar itself (yellow dashed central circle), and a circle of 1' radius (white dashed line), also centered on the Vela pulsar, but encompassing the full emission from the Vela pulsar and the Vela PWN in the same region covered by the \textit{NICER} XTI instrument. Furthermore, to obtain an estimate of the contribution from the PWN emission to the region of the Vela pulsar itself, we also extracted spectra of two circular regions of 2.5'' radius located in the innermost parts of the PWN, close to the Vela Pulsar (see dashed yellow circles in Fig.~\ref{fig: chandra_image}). The extracted spectra are presented in Fig.~\ref{fig: chandra_spectrum}. Red and green spectra correspond to the inner PWN regions, the blue spectrum corresponds to the 2.5'' region centered on the Vela Pulsar (including emission from both the Vela PSR and the inner PWN), and finally, the black spectrum was obtained from the full 1-arcmin circle encircling the Vela PSR and the entire PWN shown in Fig.~\ref{fig: chandra_image}, which approximates the emission detected in the \textit{NICER} FoV. 

We fit all these spectra simultaneously with different model variants to represent the relative contributions of the Vela PSR itself, the inner PWN, and the large (averaged) PWN, which enter into each of the selected spectral regions. The spectral components used to model each particular spectrum are indicated in Table \ref{tab:chandra_spectra} using ticks (or crosses when the components do not contribute to a certain spectrum or region), as we now describe. We model the inner PWN with a \texttt{powerlaw$_{\rm pwn}$} component, and the 2.5'' region centered on the Vela PSR, with a 3-component model including both the PSR emission (\texttt{bbodyrad$_{\rm PSR}$} + \texttt{powerlaw$_{\rm PSR}$}) and the emission from the inner PWN (\texttt{powerlaw$_{\rm pwn}$}), which we tie together to the inner PWN spectrum. These components yield $kT_{\rm bb} \approx 0.126$~keV, while $\Gamma_{\rm PSR} \approx 1.85$, and $\Gamma_{\rm pwn} \approx 1.4$. Finally, the 1' PWN spectrum is modeled by a combination of a \texttt{powerlaw$_{\rm PWN}$} non-thermal component and an \texttt{apec} thermal component, plus the PSR emission (\texttt{bbodyrad$_{\rm PSR}$} + \texttt{powerlaw$_{\rm PSR}$}), which are tied to the 1-arcmin PSR spectrum model. Considering the contribution of each individual model component to each particular spectral region, the full PWN contribution yields $kT_{apec} \approx 0.21$~keV and $\Gamma_{\rm PWN} \approx 1.56$. This simultaneous best-fitting models reach a $\chi^2 = 1212.84$ for 1068 degrees of freedom, including a common interstellar absorption (\texttt{tbabs}) with a hydrogen column density of $3.7\pm1.3 \times 10^{20}$~cm$^{-2}$, using the abundances of \cite{2000ApJ...542..914W}.

\begin{figure}%[h!]
    \includegraphics[width=\linewidth, trim={0.9cm 1.4cm 3.0cm 2.5cm}, clip]{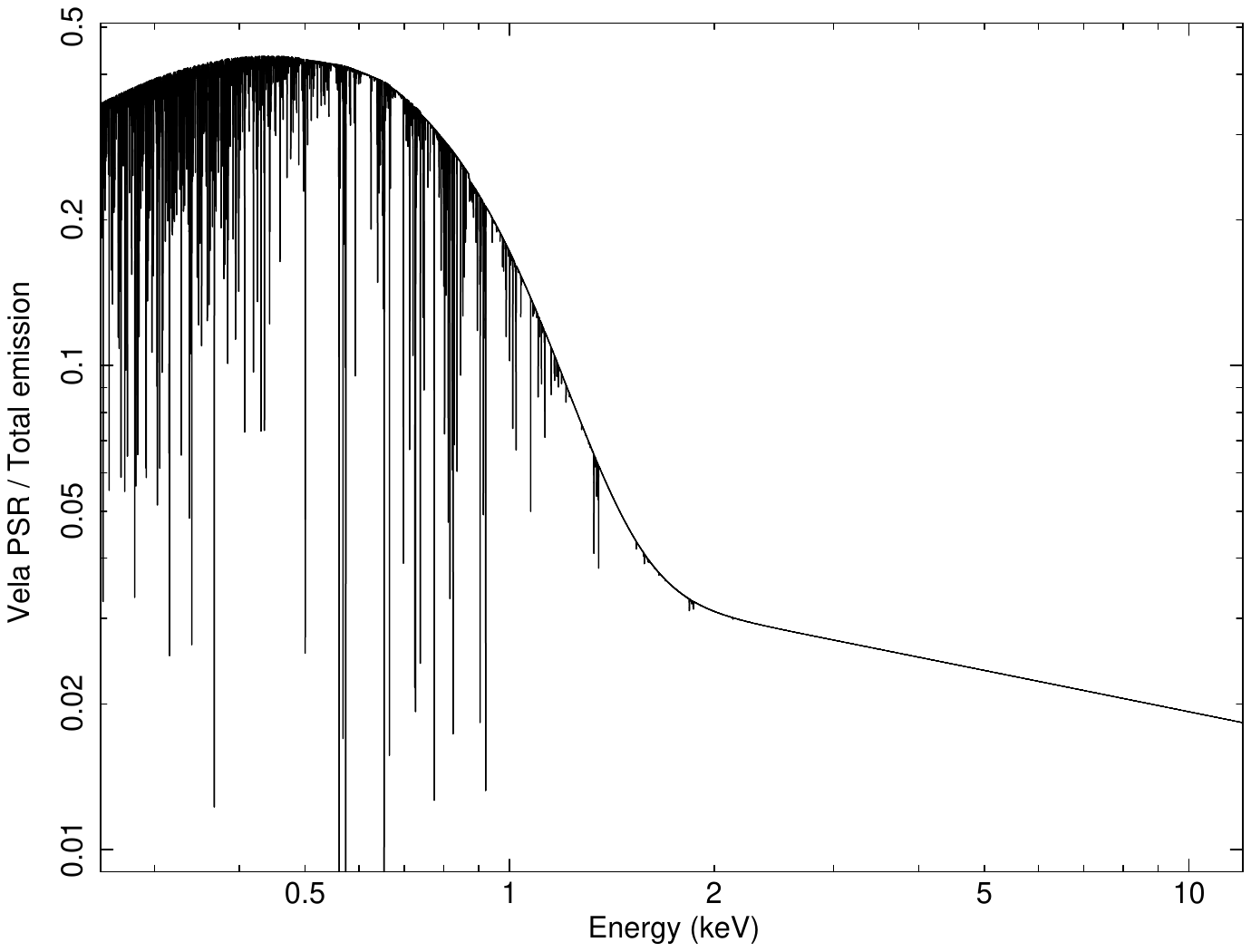}
    \caption{Vela PSR contribution to the total observed emission in the \textit{NICER} FoV as derived from the best-fitting models from \textit{Chandra}.}
    \label{fig: chandra_ratio}
\end{figure}

The best-fitting parameters to these \textit{Chandra} spectra are presented in Table~\ref{tab:chandra_spectra}, where we specifically indicate the contribution of each individual spectral component to each spectral region using either tick marks or crosses. Based on the best-fitting models, in Fig.~\ref{fig: chandra_ratio} we show the model ratio derived between the PSR contribution (\textit{bbodyrad} + \textit{powerlaw}) and the full PWN emission in the 1-arcmin circle. The figure shows that the PSR contributes to $\lesssim$40-45\% of the emission below 0.9~keV, significantly decreasing in the 0.9--2.0~keV range down to 3--4\%, and even further, down to 2\% at 10~keV, where the PWN fully dominates the detected emission in the \textit{NICER} FoV.

\begin{table*}[h]
    \centering
    \caption{Best-fitting parameters to the \textit{Chandra} spectra of the Vela PSR and PWN regions.}
    \label{tab:chandra_spectra}
    \begin{tabular}{l c c c c c}
        \hline
        Model Component & Parameter & Best-fitting value & inner PWN 2.5'' regions & PSR 2.5'' region & PWN 1' region \\
        \hline
        \texttt{tbabs} & $N_{\text{H}}~(10^{22}~{\rm cm}^{-2})$ & $0.037 \pm 0.013$ & $\checkmark$ & $\checkmark$ & $\checkmark$ \\
        \hline
        \multirow{2}{*}{\texttt{powerlaw$_{\rm pwn}$}} & $\Gamma_{\text{pwn}}$ & $1.39 \pm 0.03$ & $\checkmark$ & $\checkmark$ & $\times$ \\
         & $\text{Norm}_{\text{pwn}}^{\dagger}$ & $0.54 \pm 0.01$ & $\checkmark$ & $\checkmark$ & $\times$ \\
        \hline
        \multirow{2}{*}{\texttt{bbodyrad$_{\rm PSR}$}} & $kT_{\text{bb}}$~(keV) & $0.126 \pm 0.002$ & $\times$ & $\checkmark$ & $\checkmark$ \\
         & $\text{Norm}_{\text{bb}}^{\dagger\dagger}$ & $5300^{+750}_{-650}$ & $\times$ & $\checkmark$ & $\checkmark$ \\
        \multirow{2}{*}{\texttt{powerlaw$_{\rm PSR}$}} & $\Gamma_{\text{PSR}}$ & $1.85 \pm 0.05$ & $\times$ & $\checkmark$ & $\checkmark$ \\
         & $\text{Norm}_{\text{PSR}}^{\dagger}$ & $1.17 \pm 0.03$ & $\times$ & $\checkmark$ & $\checkmark$ \\
        \hline
        \multirow{2}{*}{\texttt{apec$_{\rm PWN}$}} & $kT_{\text{apec}}$~(keV) & $0.21 \pm 0.01$ & $\times$ & $\times$ & $\checkmark$ \\
         & $\text{Norm}_{\text{apec}}^{\dagger\dagger\dagger}$ & ($5.7 \pm 1.3) \times 10^{-3}$ & $\times$ & $\times$ & $\checkmark$ \\
        \multirow{2}{*}{\texttt{powerlaw$_{\rm PWN}$}} & $\Gamma_{\text{PWN}}$ & $1.56 \pm 0.01$ & $\times$ & $\times$ & $\checkmark$ \\
         & $\text{Norm}_{\text{PWN}}^{\dagger}$ & $46.47 \pm 0.16$ & $\times$ & $\times$ & $\checkmark$ \\
        \hline
        \multicolumn{2}{l}{$\chi^2$ (dof or bins)} & $1212.84$ (1068 dof) & $302.76$ (266 bins) & $361.03$ (302 bins) & $549.05$ (511 bins) \\
        \hline
    \end{tabular}
    \begin{flushleft}
    $^{\dagger}$ Units represent unabsorbed $\text{Flux}_{\text{2-8 keV}}$~(erg~cm$^{-2}$~s$^{-1}$).\\
    $^{\dagger\dagger}$ Units expressed in terms of $R^2/D^2$, where R is given in km and D is the distance in units of 10 kpc.\\
    $^{\dagger\dagger\dagger}$ APEC units are given in terms of $\frac{10^{-14}}{4\pi D^2} \int n_e n_H dV$.
    \end{flushleft}    
\end{table*}

\subsection{Radio timing and the 2021 Vela glitch}

On 23 July 2021, \cite{2021ATel14806....1S} reported that a giant glitch occurred in the Vela pulsar between MJD 59416.76 and MJD 59417.65. The first step to obtain accurate ephemeris for the pulsar valid for the year 2021 is a thorough characterization of the glitch. This was previously analyzed in \cite{2023MNRAS.521.4504Z}, where we reported a relative increase in frequency of $\Delta\nu/\nu \sim 1.25\times10^{-6}$ and two recovery terms, the shorter of $\sim1~\mathrm{d}$ and the longer of $\sim6.4~\mathrm{d}$, with a fractional recovery of 0.7\% and 0.2\%, respectively.

To obtain the timing solution, we first selected a $90~\mathrm{d}$ window around the glitch following the procedure of \cite{2023MNRAS.521.4504Z}, and fitted the pre-glitch TOAs to Eq.~\ref{eq:timing-model}. Then, we fitted Eq.~\ref{eq:glitch-model} to the post-glitch TOAs, including the recovery terms. We kept the parameters of the glitch frozen during the next steps in the acquisition of radio ephemeris.

\begin{figure}%[h!]
    \includegraphics[width=\linewidth]{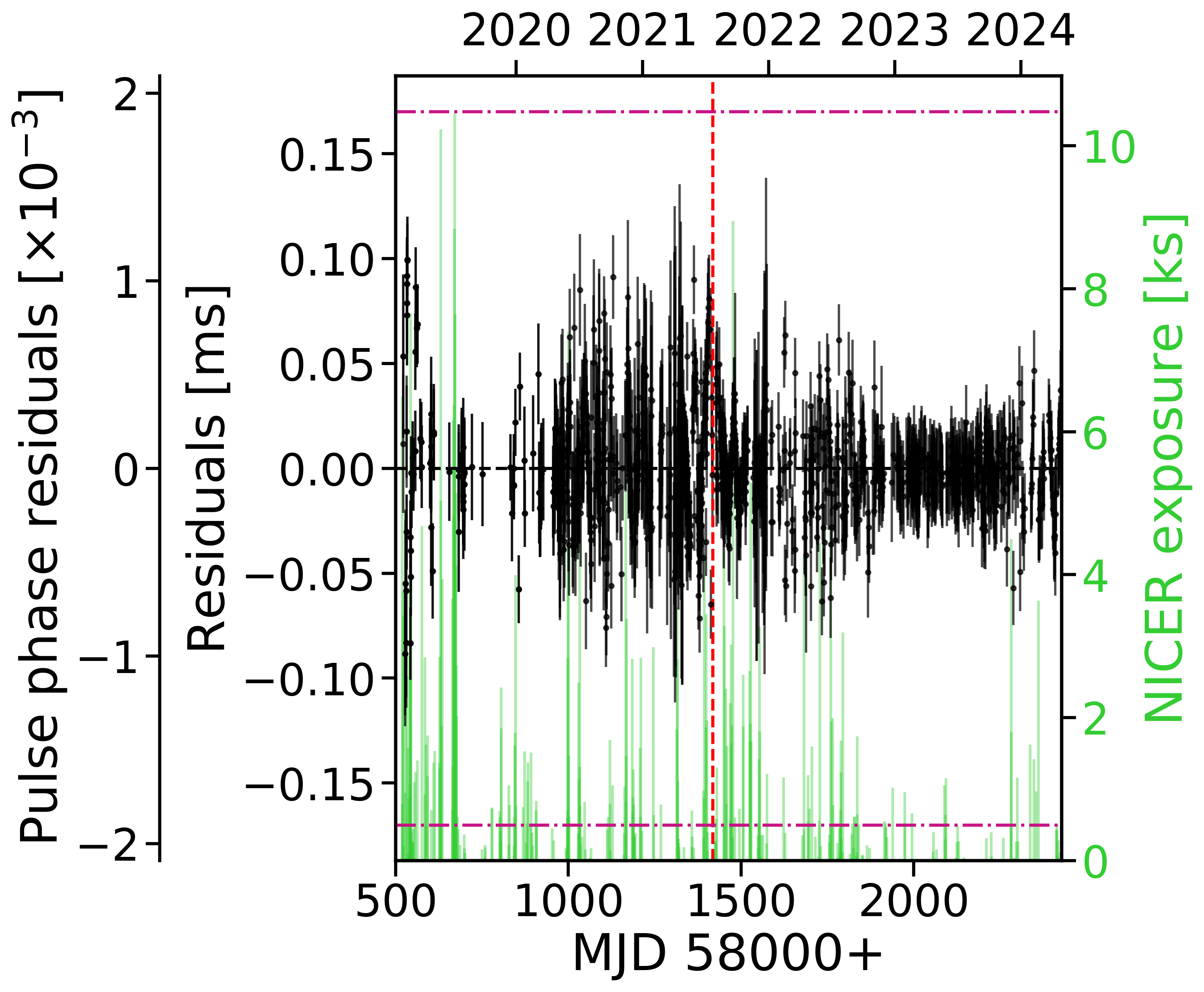}
    \caption{Residuals of the radio observations using the ephemeris obtained with IAR observations. Vertical green lines indicate the dates of available \textit{NICER} observations and their exposure. The red dashed vertical line indicates the 2021 glitch epoch. Horizontal dot-dashed lines indicate the maximum allowable residual limit to guarantee phase alignment when folding NICER observations based on the size of phase bins from its X-ray pulse profile. As can be seen in the figure, the residuals are well below this threshold.}
    \label{fig: comparison}
\end{figure}

Considering that the Vela pulsar period is $\sim89~\mathrm{ms}$ and that in the characterization of the X-ray folded profile we used 256 bins of width $\approx$0.34~ms, we needed to characterize the rotation with a timing precision of at least $0.17~\mathrm{ms}$. Therefore, the TOAs residuals fitted to Eq.~\ref{eq:timing-model} had to be below that value. With the parameters of Eq.~\ref{eq:glitch-model} frozen, we considered Eq.~\ref{eq:timing-model} with the complete set of TOAs. We then added 40 harmonics terms, as explained in \cite{2004MNRAS.353.1311H}, to remove the effects of red noise and obtained the whitened residuals shown in Fig.~\ref{fig: comparison}. The simple timing model is given in Table~\ref{tab:timing} while the fully whitened solution can be found in our repository\footnote{\url{https://github.com/ezubieta/Phase_resolved-spectroscopy-Vela-X_ray}}. The horizontal dotted-dashed lines in Fig.~\ref{fig: comparison} show that the dispersion of the residuals is strictly below the required limit to ensure accurate phase alignment of the X-ray data.

\begin{table}
  \centering    
  \caption{Parameters of the timing model valid for the data span of interest.}
   \begin{tabular}{ll}        
     \hline
     \hline
     Parameter & Value \\
     \hline
     $t_0$ (MJD)& 60408 \\
     $\nu\ (\mathrm{s^{-1}})$&  11.1842085102(2)\\
     $\nu_1\ (\mathrm{s^{-2}})$ &  $-1.57192(1)\times 10^{-11}$\\
     $\nu_2\ (\mathrm{s^{-3}})$ &  $-2.67(2)\times 10^{-21}$\\
     \hline
     $\mathrm{DM}\ (\mathrm{cm^{-3} pc})$ & 67.99(1)\\
     \hline
     $t_\mathrm{g}$ (MJD) & 59417.6194(2)\\
     $\Delta\nu_\mathrm{p}$ (s$^{-1}$) &1.38316(2) $\times 10^{-5}$\\
     $\Delta\dot{\nu}_\mathrm{p}$ (s$^{-2}$)&$-8.35(6)\times 10^{-14}$\\
     $\Delta\ddot{\nu}$ (s$^{-3}$)&$7.8(3)\times 10^{-21}$\\
     $\Delta\nu_\mathrm{d1}$\, (s$^{-1}$)&$3.15(12)\times 10^{-8}$\\
     $\tau_\mathrm{d1}$ (days) & 6.400(2)\\
     $\Delta\nu_\mathrm{d2}$\, (s$^{-1}$)&$9.9(6)\times 10^{-8}$\\
     $\tau_\mathrm{d2}$ (days) & 0.994(8)\\
     $\Delta \phi$ & $\sim0$\\
     \hline
     $\mathrm{TNGlobalEF}$ & 5.95 \\
     $\mathrm{TNGlobalEQ}$ & $-$4.82 \\
     \hline
   \end{tabular}
  \label{tab:timing}
  \tablefoot{We denote with $\nu_i$ the $i$-th derivative of $\nu$. The parenthesis denote the uncertainty in the last digit.}
 \end{table}

\subsection{X-ray profile before and after the glitch}\label{sec: prepostglitch}

\begin{figure}[h!]
    \includegraphics[width=\linewidth,trim={0.68cm 0.5cm 0.5cm 0.6cm}, clip]{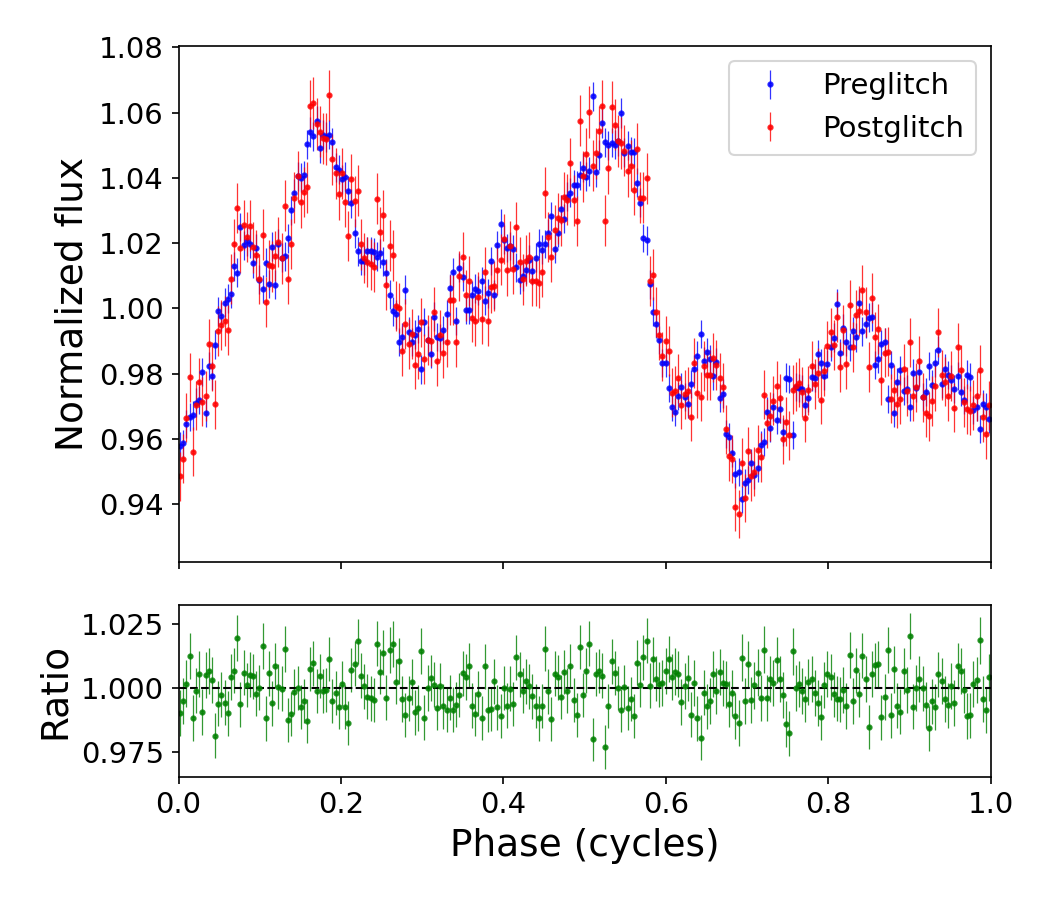}
    \caption{\textit{Top}: X-ray pulse profile obtained from the pre-glitch and post-glitch merged observations. \textit{Bottom}: Ratio between the pulse profiles.}
    \label{fig: prepostglitch}
\end{figure}

We found that the optimal energy range for the detection of the pulsed emission is 0.25--0.9~keV. From now on, we consider it as a soft energy range. Using this energy range, we first combined all observations used in this work that were performed before the 2021 glitch. We repeated the procedure with observations after the glitch. In Fig.~\ref{fig: prepostglitch} we show the resulting pre- and post-glitch profiles with 256 bins and the ratio between them. We calculated $\chi^2$ of the ratio between the two pulse profiles with respect to unity, obtaining 228.43 with 255 degrees of freedom, indicating that the pre-glitch and post-glitch folded profiles are consistent.

\subsection{Folded X-ray profile}\label{sec: folded}

\begin{figure}%[h!]
    \includegraphics[width=\linewidth, trim={0.4cm 0.5cm 0.2cm 0cm, clip}]{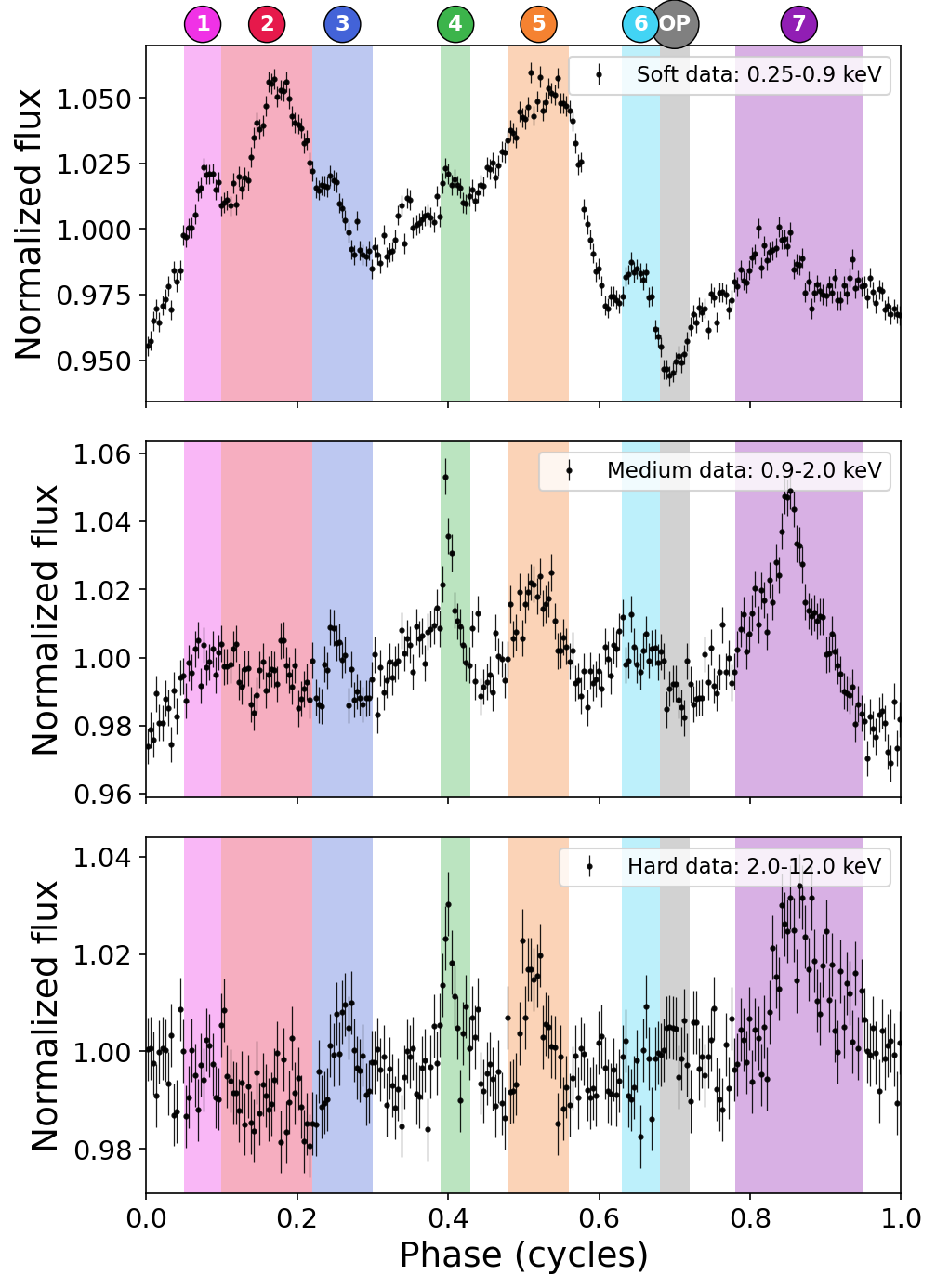}
    \caption{Pulse profile of the combined folded observations in the soft (top), medium (center), and hard (bottom) energy ranges. 
    OP is the region selected as background.}
    \label{fig: fit_6gauss}
\end{figure}

Having checked from Fig.~\ref{fig: prepostglitch} that the glitch did not affect the X-ray profile, we merged all the observations to obtain one folded profile with the best possible signal-to-noise ratio. We defined three energy ranges. The soft (optimal) range: $0.25$--$0.9~\mathrm{keV}$; a medium energy range: $0.9$--$2~\mathrm{keV}$ and a hard one in the range $2$--$12~\mathrm{keV}$. The folded profiles with 256 bins for these three energy ranges are shown in Fig.~\ref{fig: fit_6gauss}. The contribution of the pulsed emission to the soft energy range is clearly seen with $\sim$10\% of signal above the constant background. It drops to 8–9\% in the 0.9–2.0 keV range and further down to $\sim$6\% above 2 keV. The background, corresponding to the PWN and the neutron star surface, cannot be removed from \textit{NICER} observations given its $5~\mathrm{arcmin}$ resolution.  
We defined seven regions of interest that show distinct behavior. The selected regions to study their spectra are presented in Table~\ref{tab:phase_regions}. Fig.~\ref{fig: fit_6gauss} clearly shows that there are different behaviors beneath the selected regions. First, the peaks of regions 1, 3, 4 and 6 are much thinner than the rest. However, a strong contrast is observed between them:  component 4 is predominant in the medium and hard pulse profiles, while it is barely visible in the soft range. Regions 1 and 6 present softer emission, given that they are better detectable in the soft range, and they turn off for the hard range. Region 3, particularly, remains similar for the three energy ranges.
With respect to the regions 2, 5, and 7, they have wider peaks. The peak of region 7 is predominant in the medium energy range, but it is also strongly detectable in the hard energy range. The peaks of regions 2 and 5 are predominantly soft; however, the emission of region 5 is still detectable in the soft range, while the emission of region 2 is already non-detectable in the medium energy range. We also defined an off-pulse (OP) region to use as background for studying the spectra of the pulsed regions.

\begin{table}
    \centering
    \caption{Phase intervals used for the phase-resolved spectral analysis.}
    \begin{tabular}{cc}
    \hline\hline
    Region & Phase interval \\
    \hline
    OP & 0.68--0.72 \\
    R1 & 0.05--0.10 \\
    R2 & 0.10--0.22 \\
    R3 & 0.22--0.30 \\
    R4 & 0.37--0.41 \\
    R5 & 0.48--0.56 \\
    R6 & 0.63--0.68 \\
    R7 & 0.78--0.95 \\
    \hline
    \end{tabular}
    \label{tab:phase_regions}
    \tablefoot{OP denotes the off-pulse interval, used as background for the spectral extraction of regions R1--R7.}
\end{table}

\subsection{Phase-resolved X-ray spectra}

In Fig.~\ref{fig: fit_6gauss}, it is clear that some of the components persist at higher energies, while others vanish. In order to study the behavior of the different hotspots, we extracted the spectrum of the different components and the spectrum of the off-pulse (OP) region from the combination of all the observations. The regions that were used for the phase-resolved spectrum are shown in Fig.~\ref{fig: fit_6gauss}. %In particular, the widths of the hotspots that we studied were defined by the $\sigma_{i}$ of the Gaussian components shown in Table~\ref{tab:parameters}.
For fitting the spectra, we fixed the ISM hydrogen column to the value obtained with {\em Chandra} in Table \ref{tab:chandra_spectra}. We fit each region spectrum using the OP spectrum as background. For each region, we fit four different models using \textsc{cstat}: \textsc{bbodyrad}, \textsc{powerlaw}, \textsc{bbodyrad+powerlaw} and \textsc{bbodyrad+bbodyrad}. The models comparison is shown in Table \ref{tab:model_comparison}. 

For all regions, the best model is \textsc{bbodyrad} or \textsc{bbodyrad+bbodyrad}, which indicates that the pulsed emission is composed of pure thermal emission. The results show that the regions that are well detected in the hard profile of Fig.~\ref{fig: fit_6gauss}, that is, regions 4, 5 and 7, need a second small blackbody to be described. On the other side, softer regions are well described with only a single blackbody emission.

\begin{table}
\centering
\caption{Comparison of spectral models (\texttt{tbabs}$\times$\{model\}) for each phase-resolved region.}
\label{tab:model_comparison}
\begin{tabular}{clccc}
\hline\hline
Region & Model & C-stat & dof & C-stat/dof \\
\hline
\multirow{4}{*}{R1} & BB     & 1177.11 & 1172 & 1.004 \\
 & PL     & 1388.86 & 1172 & 1.185 \\
 & BB+BB  & 1177.05 & 1170 & 1.006 \\
 & BB+PL  & 1177.71 & 1170 & 1.007 \\
\hline
\multirow{4}{*}{R2} & BB     & 1135.15 & 1172 & 0.969 \\
 & PL     & 1547.72 & 1172 & 1.321 \\
 & BB+BB  & 1135.10 & 1170 & 0.970 \\
 & BB+PL  & 1547.72 & 1170 & 1.323 \\
\hline
\multirow{4}{*}{R3} & BB     & 1198.99 & 1172 & 1.023 \\
 & PL     & 1353.58 & 1172 & 1.155 \\
 & BB+BB  & 1198.73 & 1170 & 1.025 \\
 & BB+PL  & 1353.58 & 1170 & 1.157 \\
\hline
\multirow{4}{*}{R4} & BB     & 1262.50 & 1172 & 1.077 \\
 & PL     & 1223.12 & 1172 & 1.044 \\
 & BB+BB  & 1181.34 & 1170 & 1.010 \\
 & BB+PL  & 1223.76 & 1170 & 1.046 \\
\hline
\multirow{4}{*}{R5} & BB     & 1205.28 & 1172 & 1.028 \\
 & PL     & 1348.40 & 1172 & 1.151 \\
 & BB+BB  & 1127.37 & 1170 & 0.964 \\
 & BB+PL  & 1361.05 & 1170 & 1.163 \\
\hline
\multirow{4}{*}{R6} & BB     & 1132.14 & 1172 & 0.966 \\
 & PL     & 1192.47 & 1172 & 1.017 \\
 & BB+BB  & 1131.59 & 1170 & 0.967 \\
 & BB+PL  & 1131.91 & 1170 & 0.967 \\
\hline
\multirow{4}{*}{R7} & BB     & 1336.44 & 1172 & 1.140 \\
 & PL     & 1242.59 & 1172 & 1.060 \\
 & BB+BB  & 1230.58 & 1170 & 1.052 \\
 & BB+PL  & 1242.59 & 1170 & 1.062 \\
\hline
\end{tabular}
\tablefoot{$N_H$ is fixed at $0.037\times10^{22}\,\mathrm{cm^{-2}}$. BB, PL, BB+BB, and BB+PL denote \texttt{bbodyrad}, \texttt{powerlaw}, \texttt{bbodyrad+bbodyrad}, and \texttt{bbodyrad+powerlaw}, respectively. C-stat/dof is the reduced fit statistic.}
\end{table}

\begin{figure}%[h!]
    \includegraphics[width=\linewidth]{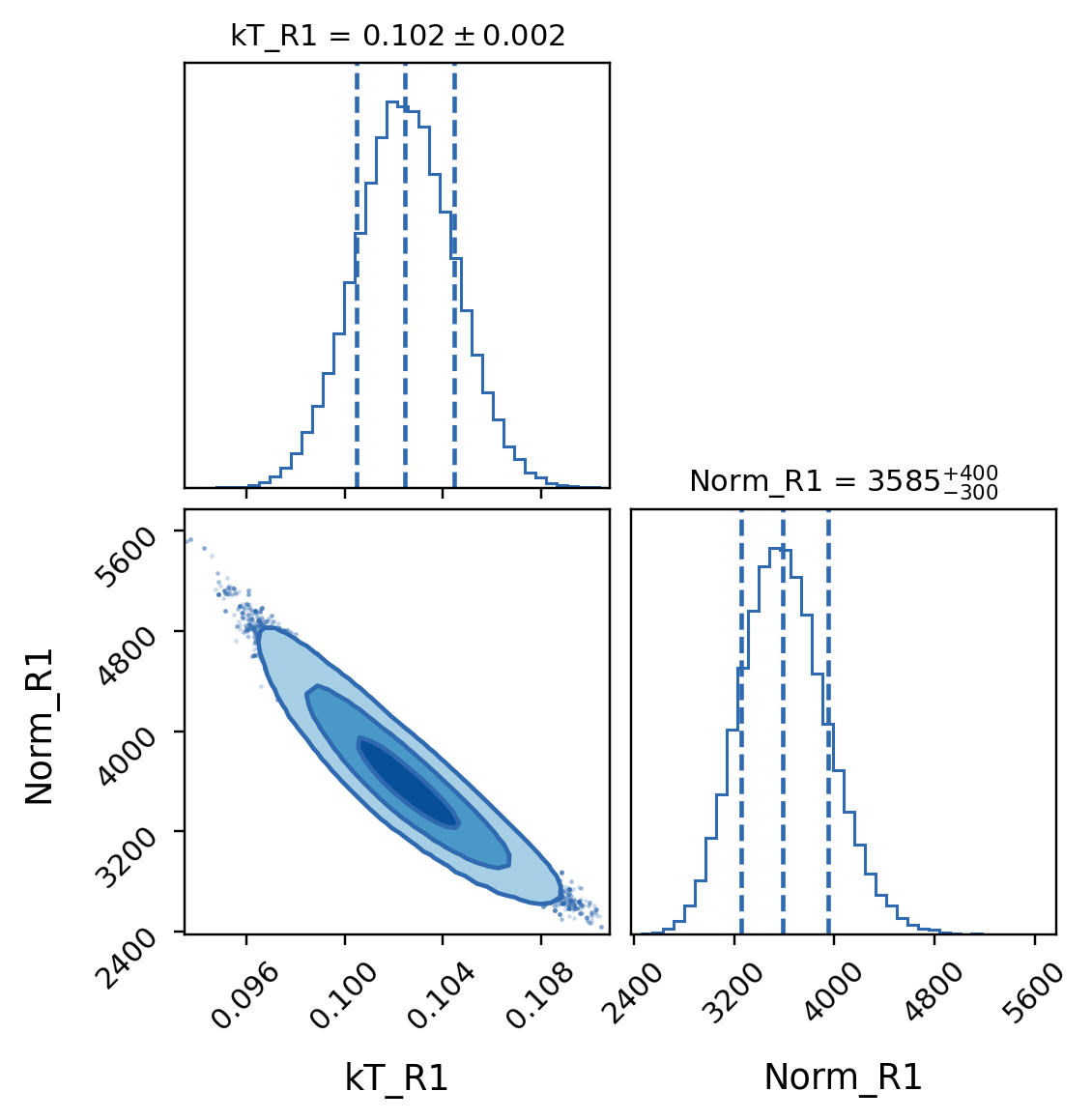}
    \caption{Cornerplot of the parameters of the region 1. $kT$ and Norm correspond to the blackbody components.}
    \label{fig: cornerplot}
\end{figure}

For each region, we fit the best model (with OP as the background)  with the \textsc{tbabs} frozen with an MCMC of length=400000, burn=200000 and 200 walkers. The posterior distribution for region 1 is shown in Fig.~\ref{fig: cornerplot}, while the posterior distribution of the rest of the regions can be seen in the repository \footnote{\url{https://github.com/ezubieta/Phase_resolved-spectroscopy-Vela-X_ray}}.

\renewcommand{\arraystretch}{1.4}
\begin{table}[h!]
  \centering
  \caption{Parameters of the phase-resolved spectroscopy (\texttt{tbabs}$\times$\texttt{bbodyrad} or \texttt{tbabs}$\times$(\texttt{bbodyrad+bbodyrad})).}
   \begin{tabular}{ll}
     \hline
     \hline
     Parameter & Value \\
     \hline
     $N_H~(10^{22}~{\rm cm}^{-2})$ & $0.037$ (fixed) \\
     $kT_{1}~(\mathrm{keV})$ & $0.102\pm0.002$ \\
     Norm$_{1}$ ($\mathrm{km}^{2}~\mathrm{kpc}^{-2}$) & $3600^{+400}_{-300}$ \\
     $kT_{2}~(\mathrm{keV})$ & $0.093\pm0.001$ \\
     Norm$_{2}$ ($\mathrm{km}^{2}~\mathrm{kpc}^{-2}$) & $7700\pm500$ \\
     $kT_{3}~(\mathrm{keV})$ & $0.086\pm0.002$ \\
     Norm$_{3}$ ($\mathrm{km}^{2}~\mathrm{kpc}^{-2}$) & $7900^{+900}_{-800}$ \\
     $kT_{4}^{(1)}~(\mathrm{keV})$ & $0.094\pm0.003$ \\
     Norm$_{4}^{(1)}$ ($\mathrm{km}^{2}~\mathrm{kpc}^{-2}$) & $4700^{+800}_{-600}$ \\
     $kT_{4}^{(2)}~(\mathrm{keV})$ & $0.33^{+0.04}_{-0.03}$ \\
     Norm$_{4}^{(2)}$ ($\mathrm{km}^{2}~\mathrm{kpc}^{-2}$) & $8^{+4}_{-3}$ \\
     $kT_{5}^{(1)}~(\mathrm{keV})$ & $0.068\pm0.006$ \\
     Norm$_{5}^{(1)}$ ($\mathrm{km}^{2}~\mathrm{kpc}^{-2}$) & $19000^{+8000}_{-5000}$ \\
     $kT_{5}^{(2)}~(\mathrm{keV})$ & $0.139^{+0.007}_{-0.006}$ \\
     Norm$_{5}^{(2)}$ ($\mathrm{km}^{2}~\mathrm{kpc}^{-2}$) & $900\pm300$ \\
     $kT_{6}~(\mathrm{keV})$ & $0.150\pm0.006$ \\
     Norm$_{6}$ ($\mathrm{km}^{2}~\mathrm{kpc}^{-2}$) & $280^{+60}_{-50}$ \\
     $kT_{7}^{(1)}~(\mathrm{keV})$ & $0.084\pm0.004$ \\
     Norm$_{7}^{(1)}$ ($\mathrm{km}^{2}~\mathrm{kpc}^{-2}$) & $4700^{+1300}_{-900}$ \\
     $kT_{7}^{(2)}~(\mathrm{keV})$ & $0.41^{+0.07}_{-0.05}$ \\
     Norm$_{7}^{(2)}$ ($\mathrm{km}^{2}~\mathrm{kpc}^{-2}$) & $3^{+2}_{-1}$ \\
     \hline
   \end{tabular}
  \label{tab:parameters-spectrum}
  \tablefoot{The subindex indicates the region to which each parameter corresponds. $N_H$ is the absorption column density, fixed at the value obtained in Table~\ref{tab:chandra_spectra} for all seven spectra. $kT_\mathrm{i}$ and Norm$_\mathrm{i}$ are the blackbody temperature and normalization of region $i$. The superscripts indicate first or second blackbody for regions modeled with two blackbodies. The uncertainties were derived from the $68.27\%$ credible interval obtained through the MCMC fit.}
 \end{table}
\renewcommand{\arraystretch}{1.0}
\begin{figure*}%[h!]
    \includegraphics[width=0.95\linewidth]{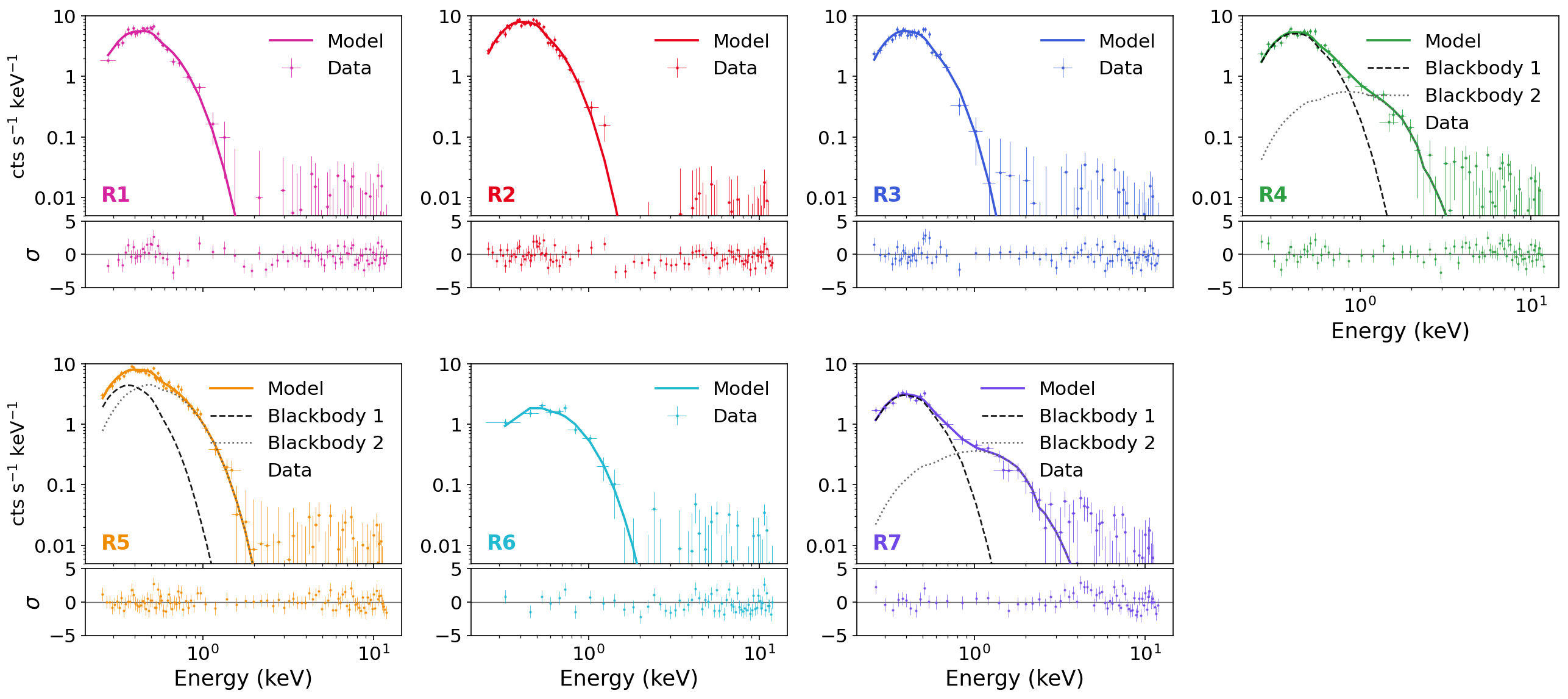}
    \caption{\textit{NICER} phase-resolved spectra of the studied regions. $kT_i$ refers to the pulsed blackbody component of each region. When the model includes two blackbodies, the components are shown with dashed and dotted lines respectively, while the total model is shown with solid lines. The data has been rebinned for visualization purposes.}
    \label{fig: plot-spectra}
\end{figure*}

In Table~\ref{tab:parameters-spectrum} we show the parameters of the phase-resolved spectroscopy with the uncertainties derived from the $68.27\%$ confidence interval of the MCMC. The respective spectra and their fits are shown in Fig.~\ref{fig: plot-spectra}. For visualization purposes only, the spectra were rebinned to a minimum of 30 counts per bin.

With respect to regions 1, 2, 3 and 6, they are well described by a single blackbody with similar temperatures, around $kT \sim 0.1\,\mathrm{keV}$. Particularly, their blackbody temperatures, $kT_1 = 0.102(2)\,\mathrm{keV}$, $kT_2 = 0.093(1)\,\mathrm{keV}$, $kT_3 = 0.086(2)\,\mathrm{keV}$, and $kT_6 = 0.150(6)\,\mathrm{keV}$, correspond to temperatures of $T_1 = 1.18(2) \times 10^{6}\,\mathrm{K}$, $T_2 = 1.08(1) \times 10^{6}\,\mathrm{K}$, $T_3 = 1.00(2) \times 10^{6}\,\mathrm{K}$, and $T_6 = 1.74(7) \times 10^{6}\,\mathrm{K}$, respectively. Furthermore, adopting a distance to the Vela pulsar of $D = 287^{+19}_{-17}\,\mathrm{pc}$~\citep{2003ApJ...596.1137D}, propagated into the size estimates via standard error propagation, the characteristic sizes of these regions are $R_1 = 1.7(1)\,\mathrm{km}$, $R_2 = 2.5(2)\,\mathrm{km}$, $R_3 = 2.5(2)\,\mathrm{km}$, and $R_6 = 0.47^{+0.06}_{-0.05}\,\mathrm{km}$.

Regions 4, 5 and 7, on the other hand, require two blackbody components to be well described. In the three cases, the main larger blackbody component also has temperatures $kT_4^{(1)} = 0.094(3)\,\mathrm{keV}$, $kT_5^{(1)} = 0.068(6)\,\mathrm{keV}$, and $kT_7^{(1)} = 0.084(4)\,\mathrm{keV}$, corresponding to $T_4^{(1)} = 1.09(3) \times 10^{6}\,\mathrm{K}$, $T_5^{(1)} = 0.79(7) \times 10^{6}\,\mathrm{K}$, and $T_7^{(1)} = 0.97(5) \times 10^{6}\,\mathrm{K}$, respectively, with characteristic sizes $R_4^{(1)} = 1.9(2)\,\mathrm{km}$, $R_5^{(1)} = 3.9^{+0.9}_{-0.6}\,\mathrm{km}$, and $R_7^{(1)} = 1.9^{+0.3}_{-0.2}\,\mathrm{km}$.

A second, smaller but substantially hotter blackbody component is also required in these three regions, with higher temperatures. In particular, $kT_4^{(2)} = 0.33^{+0.04}_{-0.03}\,\mathrm{keV}$, $kT_5^{(2)} = 0.139^{+0.007}_{-0.006}\,\mathrm{keV}$, and $kT_7^{(2)} = 0.41^{+0.07}_{-0.05}\,\mathrm{keV}$, corresponding to $T_4^{(2)} = 3.8^{+0.5}_{-0.3} \times 10^{6}\,\mathrm{K}$, $T_5^{(2)} = 1.61^{+0.08}_{-0.07} \times 10^{6}\,\mathrm{K}$, and $T_7^{(2)} = 4.8^{+0.8}_{-0.6} \times 10^{6}\,\mathrm{K}$, respectively. The corresponding emitting areas are much smaller, with $R_4^{(2)} = 0.08(2)\,\mathrm{km}$, $R_5^{(2)} = 0.8^{+0.2}_{-0.1}\,\mathrm{km}$, and $R_7^{(2)} = 0.048^{+0.016}_{-0.009}\,\mathrm{km}$.

\begin{figure}
  \centering
  \includegraphics[width=\linewidth]{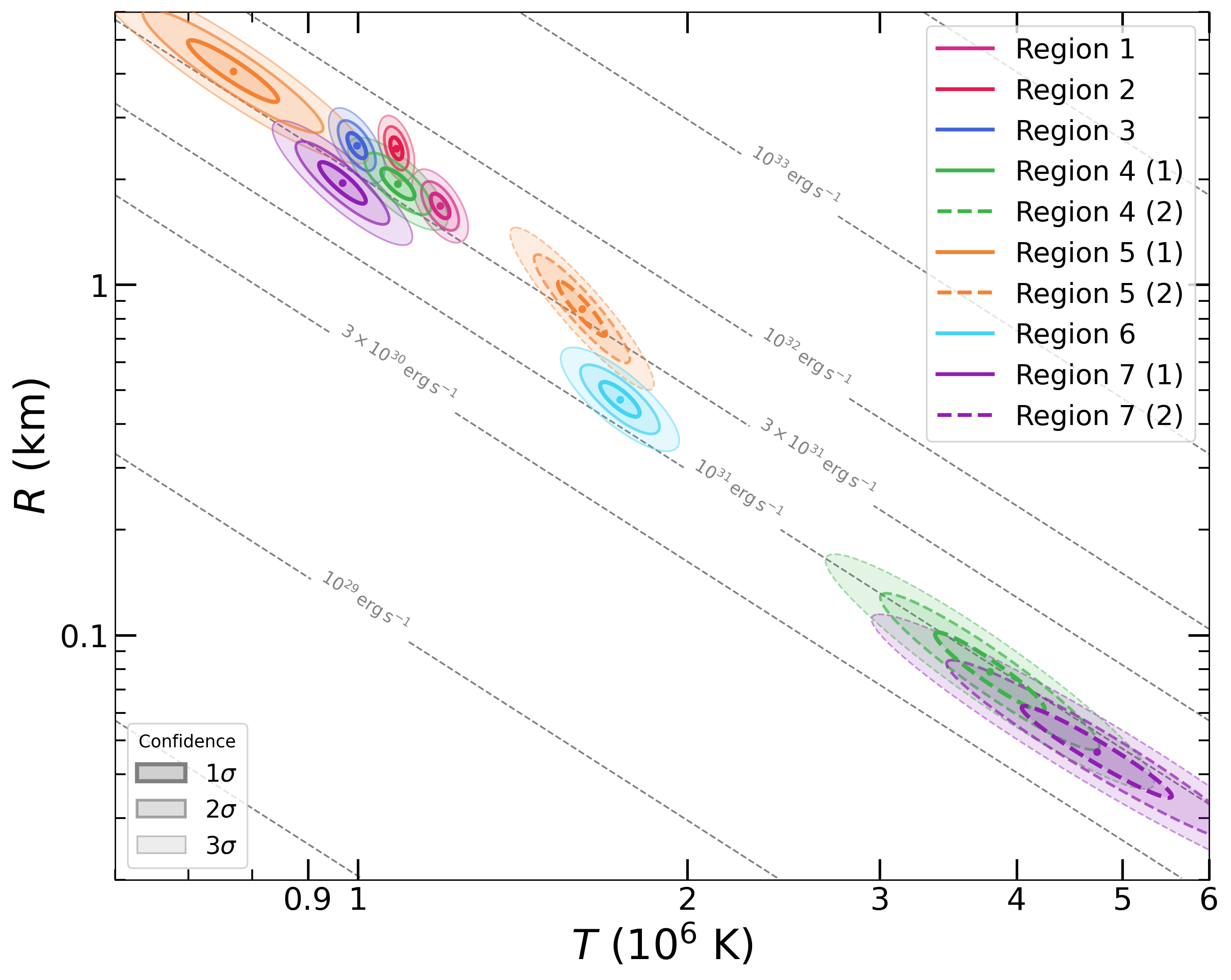}
  \caption{Relation between radius and temperature of the hotspots. The dashed lines indicate curves of constant luminosity. The displayed distributions correspond to the 1, 2 and 3 $\sigma$ regions. For regions modeled with two blackbodies, the main component is shown with solid lines and the hotter secondary component with dashed lines. %Main blackbodies are represented with solid lines while secondary hotter blackbodies of regions modeled with two blackbodies are represented with dashed lines.
  }
  \label{fig: t_vs_r}
\end{figure}

For better visualization, the $1, 2$ and $3\sigma$ distribution of the radius as a function of blackbody temperature, obtained from the posteriors of the MCMC are shown in Fig.~\ref{fig: t_vs_r}. The correlation between radius and temperature follows the Stefan--Boltzmann law, which is indicated by the dashed lines.

\section{Discussion}\label{sec: discussion}

Despite being one of the brightest and most studied pulsars, the Vela pulse profile in the soft X-ray band has not been characterized with the precision that \textit{NICER}'s sensitivity and timing capabilities allow given the lack of an accurate timing ephemeris. In addition, the 2021 giant glitch offers a rare opportunity to investigate whether such a dramatic event can alter the configuration of the soft X-ray pulse profile, a question that is not commonly studied in the soft X-ray band.

The combination of precise timing and high-cadence X-ray monitoring of the Vela pulsar by NICER makes this an optimal dataset to tackle both questions. We obtained precise ephemeris of the pulsar since 2019 using high-cadence observations from the IAR, spanning 29 months before the 2021 glitch and 33 months following it until the 2024 glitch. These ephemeris were then used to fold 321 \textit{NICER} observations, taking advantage of \textit{NICER}'s regular monitoring campaign of the Vela pulsar. Additionally, we analyzed the \textit{Chandra} observation closest in time to the glitch to quantify the pulsar's contribution to the total emission detected by \textit{NICER}, since \textit{NICER} cannot spatially resolve the pulsar from the surrounding PWN due to its wide field of view.

Our analysis reveals that the Vela pulse profile in the soft X-ray band is remarkably complex, exhibiting multiple peaks with distinct energy dependencies across the energy sub-bands. Furthermore, no significant changes in the profile were detected between the pre- and post-glitch epochs, suggesting that pulsar's X-ray emission geometry remained unaffected despite the 2021 giant glitch. This is consistent with the findings of \cite{2000ApJ...536..948S}, who reported no significant changes in the pulse structure after the 1996 Vela giant glitch.

We note that magnetospheric changes coincident with a glitch have been reported for other pulsars with radio observations. Particularly, a nulling pulse was found for Vela itself during its December 2016 glitch \citep{2018Natur.556..219P}. These changes are usually interpreted as evidence of glitch-induced magnetospheric reconfiguration. For the 2021 glitch analyzed here, however, a dedicated radio monitoring campaign found no notable change in Vela's individual pulse behavior \citep{2023MNRAS.521.4504Z}, in agreement with the stability of the X-ray pulse profile we report in this work. This double null result indicates that despite having undergone a giant glitch, neither its magnetic topology nor the thermal profile of its surface appear to be significantly affected by this event.

The X-ray pulse profile of the Vela pulsar has been studied across a broad energy range by multiple observatories over the past three decades, revealing a complex and strongly energy-dependent morphology. The first unambiguous detection of pulsed soft X-ray emission was reported by \cite{1993Natur.361..136O} using ROSAT, which revealed a multi-peaked profile below 1 keV with a pulsed fraction of approximately 10\%, embedded within the bright surrounding pulsar wind nebula (PWN). Subsequent ROSAT HRI monitoring by \cite{2000ApJ...536..948S} confirmed the pulse structure, detecting one broad and two narrow peaks. In addition, they did not find detectable changes in pulse morphology or X-ray luminosity following a large glitch. This early result already suggested a degree of stability of the X-ray emission geometry with respect to glitch events, albeit with limited sensitivity.
The angular resolution of ROSAT was insufficient to fully disentangle the pulsar from its surrounding PWN, a limitation that was later overcome by \textit{Chandra}. \cite{2001ApJ...552L.129P} used the \textit{Chandra} LETG and ACIS-S instruments to resolve the pulsar spectrum for the first time, revealing a two-component structure: a dominant soft thermal component below $\sim$1.8 keV, consistent with emission from the neutron star surface, and a harder non-thermal power-law component that dominates at higher energies. \cite{2007Ap&SS.308..601M} later exploited the high photon statistics of XMM-Newton to perform phase-resolved spectroscopy, confirming the multi-component nature of the pulse profile and characterizing the spectral variations with rotational phase across the soft X-ray band.

A comprehensive multiwavelength picture was assembled by \cite{2015MNRAS.449.3827K}, who compiled pulse profiles from radio through soft gamma-rays and showed that the profile structure evolves continuously with energy, with the relative contribution of thermal and non-thermal components shifting between the soft and hard X-ray sub-bands. The multiwavelength pulse profiles compiled by \cite{2025ApJ...988..200L} using \textit{NICER} (0.3–0.8 keV and 0.8–2 keV), \textit{NuSTAR}, and \textit{Fermi}-GBM further underscore how profoundly the profile morphology changes across energy bands, with the softest \textit{NICER} band dominated by thermal surface emission while the harder bands increasingly reflect the non-thermal magnetospheric contribution.

Building on this multiwavelength context, we present a detailed characterization of the Vela pulse profile in the soft X-ray band using the \textit{NICER} dataset. Given that \textit{NICER} cannot spatially resolve the pulsar from the surrounding PWN, we first quantify the pulsar's contribution to the total detected emission by leveraging the superior angular resolution of the contemporaneous \textit{Chandra} observation. With the pulsar fraction established, we proceed with a detailed analysis of the energy-resolved pulse profile across the 0.25–12 keV band, decomposing its multiple peaks and characterizing the spectrum of each component individually.

The Vela pulsar is deeply embedded within its pulsar wind nebula, which dominates the X-ray emission in the \textit{NICER} field of view. As shown by our \textit{Chandra} analysis, the absorption column density obtained, $N_\mathrm{H}=(0.037\pm0.013)\times10^{22}~\mathrm{cm^{-2}}$, is consistent with previous measurements \citep{2001ApJ...552L.129P}. The pulsar contribution to the total observed emission represents around 40-45\% of the flux below 0.9 keV, dropping to 3–4\% in the 0.9–2.0 keV range and further down to $\sim2$\% above 2 keV, where the PWN dominates. We find that the full PWN contribution yields $kT_{apec} \approx 0.21$~keV and $\Gamma_{\rm PWN} \approx 1.56$, similar to values previously reported with \textit{NuSTAR} observations \cite{2024ApJ...975..199K} and \textit{XMM-Newton} observations \cite{2004AdSpR..33..503M}. This strong contamination makes phase-resolved spectral studies of the Vela pulsar particularly challenging with instruments of limited angular resolution such as \textit{NICER} ($\sim$5'). Despite this difficulty, by merging 321 \textit{NICER} observations and exploiting the high-cadence radio ephemeris obtained at the IAR, we were able to construct a high signal-to-noise folded profile that reveals a rich and complex structure both in pulse phase and across energy ranges. The resulting profile shows different peaks whose relative amplitudes vary significantly with energy, reflecting the distinct spectral properties of the individual emission regions. This suggests a complex magnetic configuration that concentrates the heating in multiple regions of the surface, and that was not detectably affected by the glitch.

Regarding the pulsed emission components, we obtained well-constrained normalizations for the seven regions identified in the pulse profile. While four of these regions are described with one blackbody component, the other three were better described with two blackbodies, with the second component much hotter and smaller. The temperatures of the hotspots range between  $0.79(7) \times 10^{6}\,\mathrm{K} \leq T \leq 4.8^{+0.8}_{-0.6} \times 10^{6}\,\mathrm{K}$, while their sizes vary between $0.048^{+0.016}_{-0.009}\,\mathrm{km} \leq R \leq 3.9^{+0.9}_{-0.6}\,\mathrm{km}$. 

These hotspots are expected considering the polar-cap heating from back-flowing relativistic particles (``back-heating'') generated within the pulsar magnetosphere, which deposits additional energy over localized regions of the surface \citep{2002ApJ...568..862H}. Although individual uncertainties on the temperature and size of each hotspot are relatively large, their luminosity is tightly constrained, as illustrated in Fig.~\ref{fig: t_vs_r}. These luminosities range between $5\times10^{30}~\mathrm{erg}~\mathrm{s^{-1}} \lesssim L < 1\times10^{32}~\mathrm{erg}~\mathrm{s^{-1}}$, similar to the values obtained by \cite{2007Ap&SS.308..601M}.
In particular, Regions 5 and 7 of this work correspond to Peaks 1 and 2 of \cite{2025ApJ...988..200L}, respectively. To corroborate that they are not the extrapolation of the non-thermal component seen in gamma rays, we calculated the off-pulse fluxes for these regions with \textit{NICER}. The fluxes derived close to the peak are $E^{2}F_{E}=(1.91\pm0.06)\times10^{-6}\,\mathrm{MeV^{2}\,cm^{-2}\,s^{-1}\,MeV^{-1}}$ at 0.5\,keV for R5, and $(3.9\pm0.8)\times10^{-7}\,\mathrm{MeV^{2}\,cm^{-2}\,s^{-1}\,MeV^{-1}}$ at 1\,keV for R7. These values lay above the extrapolation of the non-thermal components of Peaks 1 and 2 of \cite{2025ApJ...988..200L}, which corroborates that the flux we detect with \textit{NICER} is dominated by thermal emission.

The results presented here highlight the value of combining high-cadence radio monitoring with X-ray observatories. The precise ephemeris provided by the IAR was essential to coherently fold the \textit{NICER} observations and resolve the fine structure of the pulse profile, a step that would not have been possible with the X-ray data alone. This approach is broadly applicable to other young, bright pulsars monitored by the IAR, for which \textit{NICER} archival data may be similarly exploited. In particular, this study can also be replicated in gamma-ray pulse profile studies. The IAR monitoring program therefore represents a natural resource for multiwavelength timing campaigns targeting southern pulsars, including sources that have not yet been studied with the sensitivity and cadence that modern X-ray and gamma-ray observatories can provide.

\begin{acknowledgements}
We thank the referee for helping us improve the results and presentation of this work. We are grateful to the IAR technical team for their unwavering support. Their work is essential to keeping the IAR radio telescopes in operation, overcoming the ongoing difficulties within the scientific community in Argentina. EZ is a CONICET Fellow. JAC and FG are CONICET researchers. EZ, JAC and FG acknowledge support from PIP 0113 (CONICET). JAC was supported by Consejería de Economía, Innovación, Ciencia y Empleo of Junta de Andalucía as research group FQM-322. JAC and FG were also supported by grant PID2022-136828NB-C42 funded by the Spanish MCIN/AEI/ 10.13039/501100011033 and “ERDF A way of making Europe”. 
\end{acknowledgements}

\bibliographystyle{aa} % style aa.bst
\bibliography{biblio} % your references biblio.bib

\end{document}